\documentstyle[12pt]{article}
\newlength{\overeqskip}
\newlength{\undereqskip}
\newcommand{\nc}{\newcommand}
\nc{\be}[1]{\begin{equation} \mbox{$\label{#1}$}}
\nc{\bea}[1]{\begin{eqnarray} \mbox{$\label{#1}$}}
\nc{\Label}[1]{\label{#1}}
\nc{\Bibitem}[1]{\bibitem{#1}}
\nc{\ba}{\begin{array}}
\nc{\ea}{\end{array}}

\def\bort#1{ }
\nc{\eea}{\vspace{\undereqskip}\end{eqnarray}}
\nc{\ee}{\vspace{\undereqskip}\end{equation}}
\nc{\nn}{\nonumber \\}
\nc{\vek}[1]{{\rm\bf #1}}

\def\pa{\partial}

\def\>{\rangle}
\def\<{\langle}

\def\inv#1{\frac{1}{#1}}
\def\simleq{\; \raise0.3ex\hbox{$<$\kern-0.75em
      \raise-1.1ex\hbox{$\sim$}}\; }
\def\simgeq{\; \raise0.3ex\hbox{$>$\kern-0.75em
      \raise-1.1ex\hbox{$\sim$}}\; }
\nc{\at}[2]{\left.#1\right|_{#2}}
\nc{\eqs}[2]{\mbox{Eqs.~(\ref{#1},\,\ref{#2})}}
\nc{\eq}[1]{\mbox{Eq.~(\ref{#1})}}
\nc{\figs}[2]{\mbox{Figs.~(\ref{#1},\,\ref{#2})}}
\nc{\fig}[1]{\mbox{Fig.~(\ref{#1})}}
\nc{\figcap}[1]{\begin{quote}\refstepcounter{figure}
        {\bf Figure \thefigure}: {\small #1}\end{quote}}
\def\lsim{\; \raise0.3ex\hbox{$<$\kern-0.75em
      \raise-1.1ex\hbox{$\sim$}}\; }
\def\gsim{\; \raise0.3ex\hbox{$>$\kern-0.75em
      \raise-1.1ex\hbox{$\sim$}}\; }

\nc{\rf}[1]{(\ref{#1})}
\nc{\slsh}{\hskip-5pt/}
\nc{\eff}{{\rm eff}}
\nc{\Tc}{T_{\rm crit}}
\nc{\om}{\omega}
\nc{\Om}{\Omega}

\def\GeV{\rm GeV}

\nc{\apj}[3]{{\it  Ap.\ J.\ }{{\bf #1} {(#2)} {#3}}}
\nc{\apjl}[3]{{\it  Ap.\ J.\ Lett.\ }{{\bf #1} {(#2)} {#3}}}
\nc{\app}[3]{{\it Astropart.\ Phys.\ }{{\bf #1} {(#2)} {#3}}}
\nc{\ijmp}[3]{{\it Int.\ J.\ Mod.\ Phys.\ }{{\bf #1} {(#2)} {#3}}}
\nc{\ijtp}[3]{{\it Int.\ J.\ Theor.\ Phys.\ }{{\bf #1} {(#2)} {#3}}}
\nc{\mpl}[3]{{\it  Mod.\ Phys.\ Lett.\ }{{\bf #1} {(#2)} {#3}}}
\nc{\ncim}[3]{{\it  Nuov.\ Cim.\ }{{\bf #1} {(#2)} {#3}}}
\nc{\np}[3]{{\it  Nucl.\ Phys.\ }{{\bf #1} {(#2)} {#3}}}
\nc{\pr}[3]{{\it Phys.\ Rev.\ }{{\bf #1} {(#2)} {#3}}}
\nc{\prl}[3]{{\it Phys\ Rev.\ Lett.\ }{{\bf #1} {(#2)} {#3}}}
\nc{\pl}[3]{{\it  Phys.\ Lett.\ }{{\bf #1} {(#2)} {#3}}}
\nc{\prep}[3]{{\it Phys\. Rep.\ }{{\bf #1} {(#2)} {#3}}}
\nc{\phys}[3]{{\it Physica\ }{{\bf #1} {(#2)} {#3}}}
\nc{\rmp}[3]{{\it  Rev.\ Mod.\ Phys.\ }{{\bf #1} {(#2)} {#3}}}
\nc{\rpp}[3]{{\it Rep.\ Prog.\ Phys.\ }{{\bf #1} {(#2)} {#3}}}
\nc{\sjnp}[3]{{\it Sov.\ J.\ Nucl.\ Phys.\ }{{\bf #1} {(#2)} {#3}}}
\nc{\spjetp}[3]{{\it Sov.\ Phys.\ JETP\ }{{\bf #1} {(#2)} {#3}}}
\nc{\yf}[3]{{\it Yad.\ Fiz.\ }{{\bf #1} {(#2)} {#3}}}
\nc{\zetp}[3]{{\it Zh.\ Eksp.\ Teor.\ Fiz.\  }{{\bf #1}  {(#2)} {#3}}}
\nc{\zp}[3]{{\it Z.\ Phys.\ }{{\bf #1} {(#2)} {#3}}}
\nc{\ibid}[3]{{\sl ibid.\ }{{\bf #1} {#2} {#3}}}
\def\ext{{\rm ext}}
\begin{document}
%
%
\begin{titlepage}
\pagestyle{empty}
\baselineskip=21pt
\rightline{HIP-1998-31/TH}
\rightline{NORDITA-98/45 HE}
\rightline{SUITP-98-10}
\rightline{hep-ph/9806403}
\rightline{June 18, 1998}
\vskip 0.2in
\begin{center}
  {\Large{\bf Strongly First Order Electroweak Phase Transition\\[2mm]
induced by Primordial Hypermagnetic Fields}}
\end{center}
\vskip 0.1in
\begin{center}
  {\large Per Elmfors}\footnote{elmfors@physto.se}  \\
  {\it Department of Physics, University of Stockholm,\\
       P.O. Box 6730, S-11385, Stockholm, Sweden
       }\\[1mm]
  {\large Kari Enqvist}\footnote{ enqvist@rock.helsinki.fi} \\
  {\it Department of Theoretical Physics and
       Helsinki Institute of Physics,\\
       P.O.~Box 9, FIN-00014 University of Helsinki, Finland
       } \\[1mm]
  {\large Kimmo Kainulainen}\footnote{kainulai@nordita.dk} \\
  {\it NORDITA, Blegdamsvej 17, DK-2100, Copenhagen \O, Denmark
        } \\

\end{center}
\vskip 0.1in
\centerline{ {\bf Abstract} }
\baselineskip=18pt
\vskip 0.1truecm
We consider the effect of the presence of a hypermagnetic field at the
electroweak phase transition. Screening of the $Z$-component inside a bubble
of the broken phase delays the phase transition and makes it stronger first
order.  We show that the sphaleron constraint can be evaded for $m_H$ up to
100~GeV if a $B_Y\gsim 0.3 T^2$ exists at the time of the EW phase transition,
thus resurrecting the possibility for baryogenesis within the minimal standard
model (provided enough $CP$ violation can be obtained).  We estimate that for
$m_H\gsim 100$~GeV the Higgs condendsate behaves like a type~II
superconductor with $Z$-vortices penetrating the bubble. Also, for such high
Higgs masses the minimum $B_Y$ field required for a strong first order phase
transition is large enough to render the $W$-field unstable
towards forming a condensate which changes the simple picture of the symmetry
breaking.
\\
\end{titlepage}
%
%
\baselineskip=20pt
\textheight8.3in\topmargin-0.0in\oddsidemargin-.0in

It is by now generally accepted that baryogenesis is not possible in the
minimal Standard Model (MSM). The reason is that the sphaleron constraint,
which is needed to ensure the survival of the baryon number against sphaleron
induced erasure, cannot be satisfied in the MSM, no matter what the mass of
the
Higgs boson is \cite{KLRSa}, because the radiative correction induced to the
effective action by the top quark makes the transition too weakly first order.
While this result was first borne out from a detailed lattice
calculation, it can actually be robustly seen already at the level of
the simple 1-loop effective potential.

However, it is conceivable that the presence of a large scale homogenous
magnetic field could change the situation. Magnetic fields are a generic
feature in the very early universe, and there has been several suggestions for
mechanisms giving rise to strong fields (for a recent review, see \cite{joel}).
For instance, certain types of inflationary models can produce magnetic fields
extending over horizon distances \cite{inflation}. These models must be able to
break the conformal invariance to escape the flux freezing constraint, which
would dictate that $B\sim R^{-2}$. Another possibility \cite{ferro} could be
that the Yang-Mills vacuum is unstable and the true ground state has a non-zero
$\vec B$, as suggested by Savvidy \cite{savvidy}. It however
appears unlikely that the original argument by Savvidy can be carried
over to a finite temperature environment \cite{oikeinko}. It has also
been proposed that a right-handed electron asymmetry, generated at the
GUT scale and conserved down to TeV-range in temperature \cite{CKO},
could give rise to a hypercharge magnetic field $B_Y$ via a
Chern-Simons term \cite{joyceshapo}.

No matter what the origin of the primordial magnetic field, it could
have interesting consequences for physics at the electroweak scale.
For example, a hypermagnetic field could have a pronounced effect on the
expansion of the phase transition bubble which would have to push
against the field lines frozen in the plasma outside the bubble.
In this letter we concentrate on another interesting issue, namely
whether a magnetic field could make the electroweak phase transition
more strongly first order by way of increasing the (Gibbs) free energy
difference between the broken and unbroken phases (this effect was recently
discussed also by Giovannini and Shaposnikov \cite{GS}). Naively the
argument goes as follows: Above the EW transition temperature any
primordial large-scale field must be aligned in the hypercharge
direction, because in the symmetric phase the transverse $SU(2)$ fields are
screened over distances larger than the inverse of the magnetic mass sacale
$m_g\sim  g^2 T$.  In the broken phase however, only the projection of the
original hypermagnetic field onto the Maxwell direction (corresponding to the
usual photon) is unscreened, while the $Z$-component becomes screened by
surface currents. Expelling the $Z$-component costs in energy, leading to a
higher Gibbs free energy in the broken phase. The situation is similar
to a superconductor, and since the mass of the Higgs field is
relatively small compared to the $Z$-mass around the phase transition,
one expects it to be typically of type~I, i.e. without vortex formation.
The $B_Y$-field is locally conserved and the expelled component has to go
outside the bubble. The expelled field can obviously have a profound effect
on the completion of the phase transition. However, we do not expect that to
be the case for the bubble nucleation problem.

Because the early universe is an excellent conductor \cite{cond}, the magnetic
flux was frozen in the plasma on large scales.  There was however some
magnetic diffusion, leading to  a straightening out of the entangled field
lines
at small scales. From the MHD equation one finds that inhomogeneities with
scales less than  $L_0$ have decayed before the onset of the EW phase
transition, with $L_0$ given by
\be{dec}
L_0\simeq (t_{EW}/\sigma)^{1/2}\simeq 2\times 10^5 \GeV^{-1}~,
\ee
where $\sigma\simeq 10T$ is the conductivity, and here we take
$T\simeq 100$GeV. A typical size of the EW bubble is expected to be roughly
of the order
$100/T-1000/T$. Therefore the bubble formation takes place in the background of
essentially constant field (apart from shortlived microscopic fluctuations), no
matter how random it was originally, and for our purposes we may assume
$B_Y^\ext\sim$ const. We parametrize
\be{para}
        B^\ext_Y=b(T)T^2~,
\ee
and study the consequences as a function of $b$.

The constraints on $b$ are uncertain and follow mainly
from the microwave background and
primordial nucleosynthesis considerations.  Assuming that the whole
anisotropy measured by COBE is due to magnetic stresses one obtains
\cite{cmblimit} $B \le 3.4\times 10^{-9} {\rm G}\;
(\Omega_{0}h_{50}^2)^{1/2}$ if the field can be taken homogeneous at
the recombination. This limit implies a superhorizon size coherence
length for $B$ and hence must refer to a mean-root-square field
$\sqrt{\langle B^2\rangle }$, which is expected to be much
smaller than the small scale field and thus does not constrain $b$
directly. The observed Helium abundance implies \cite{nucls} that
$B\lsim 10^{12}$G at $T\simeq 0.1$ MeV and at length scales greater than 10 cm.
If the flux is covariantly conserved, the nucleosynthesis limit implies that
$B\lsim T^{2}$ at scales greater than $10^{-12}L_{horizon}$.  However,
due to turbulence in the primordial magnetoplasma we also expect an
inverse cascade
\cite {beo1} of magnetic energy to take place, whereby power from small
distance
scales is transferred into larger scales. This means that the simple
scaling law
$B\sim B_0T^2/T_0^2$ may not hold and that the field at earlier times could at
small scales have been even larger than $T^2$.
Since the uncertainties are large we shall assume
that $b(T)\simleq 1$, with a slowly
varying $b$ just before the time of EW phase transition. There is however no
direct bound on the magnetic energy densities as long as it is much smaller
than the radiation energy density which only implies $b(T)\ll 10$.

Let us  first consider the sphaleron erasure constraint in the  MSM
in the absence of any external fields.  Including the radiative corrections
from all the known Standard Model particles, one obtains the effective
potential
\be{veff1loop}
V_{\rm eff}(\phi , T) \simeq -\frac{1}{2}(\mu^2 - \alpha T^2)\phi^2
  -  T \delta \phi^3 + \frac{1}{4}(\lambda - \delta\lambda_T )\phi^4~.
\ee
To the one loop order in the MSM the potential parameters are
\be{zeroT}
\mu^2 = \left(\lambda- \frac{44}{3}B\right) v^2~,   \qquad
\lambda = \frac{m^2_H}{2v^2} + \frac{32}{3}B~,
\ee
where
\be{Beq}
B \equiv \frac{1}{64\pi^2v^4}\left( 3M_Z^4 + 6M_W^4 -12m_t^4\right)~,
\ee
and
\bea{values}
\alpha &=& \frac{1}{4v^2}\left( M_Z^2 + 2M_W^2 + 2m_t^2 \right)~,
\nonumber\\
\delta &=& \frac{1}{6\pi v^3}(M^3_Z+2M^3_W)~,
\nonumber\\
\delta\lambda_T  &=&  \frac{1}{16 \pi^2 v^4}
            \left( 3 M_Z^4 f_B(M_Z,T) + 6 M_W^4 f_B(M_W,T)
     - 12 m_t^4 f_F(m_t,T) \right)~,
\eea
with
\be{ahva}
f_X(M,T) \equiv \ln \frac{M^2}{T^2} + \frac{25}{6} - c_X~,
\ee
and $c_B \simeq 5.41$ and $c_F \simeq 2.64$.

 From (\ref{veff1loop}) one can easily find that at the critical temperature,
when the two minima of the potential become degenerate
\be{constraint}
\frac{\phi}{T} = \frac{2\delta}{\lambda-\delta \lambda_T}~.
\ee
In MSM $\phi/T$ has a maximum value of 0.55 when $m_H=0$ and decreases
monotonically as $m_H$ increases. However, avoiding the
sphaleron wash-out requires that \cite{KLRSa}
\be{sphcons}
\left(\frac{\phi}{T}\right)_{\rm min} \simeq 1.0 - 1.5~,
\ee
and therefore it follows from (\ref{constraint}) that baryogenesis is not
possible in MSM.  Note that this conclusion is robust, because one expects
that for accepted values of $\phi/T \gsim 1$ in \eq{sphcons}, the effective
potential approach is quite reliable.

Let us now address the question of how the presence of a homogenous large
scale magnetic field affects the order of the transition. Consider first
the MSM with a background field at the tree level and at the zero
temperature.
We induce the coupling to the external hyper electromagnetic field by  adding
a source term for the hyper field tensor $f_{\mu\nu}$ to the Lagrangian
\be{coupling}
        {\cal L} = {\cal L}_{\rm MSM}
        +\frac{1}{2}f_{\mu\nu}f^{\mu\nu}_{\rm ext}~,
\ee
where
\be{lagMSM}
{\cal L}_{\rm MSM} = | D \Phi |^2 -
V(\Phi^\dagger\Phi) - \frac{1}{4}F^a_{\mu\nu}F^{a,\mu\nu} -
\frac{1}{4}f_{\mu\nu}f^{\mu\nu} + ...
\ee
Here $D_\mu=\pa_\mu+i\frac{g}{2}\sigma^aA^a_\mu+i\frac{g'}{2}A^Y_\mu$
is the usual covariant derivative, $F^a_{\mu\nu}$ is the $SU(2)$ gauge field
strength, and we have left out parts
corresponding to fermions.
They are not relevant here since at the EW scale there are no large
scale fermionic condensates.

The vacuum expectation values of the fields are determined from the
requirement
that the modified action be stationary with respect to variations in $\Phi$,
the $SU(2)$ gauge fields $A^a_\mu$ and the hypercharge field $A^Y_\mu$:
\bea{fieldeqTree}
  D^2\Phi -  V'(\Phi^\dagger\Phi)\Phi &=& 0~,
\nonumber \\
  (D_\nu F^{\mu\nu})^a -
\frac{ig}{2}\left( \Phi^\dagger \sigma^a D_\mu \Phi
            - (D_\mu \Phi )^\dagger \sigma^a \Phi\right) &=& 0~,
\nonumber \\
  \partial_\nu (f^{\mu\nu}-f^{\mu\nu}_{\rm ext}) - \frac{ig'}{2}
  \left( \Phi^\dagger D_\mu \Phi  - (D_\mu \Phi^\dagger) \Phi\right) &=& 0~.
\eea
One can see that there is a simple solution for these equations:
\be{unbrokenTree}
f_{\mu\nu} = f^{\mu\nu}_{\rm ext}, \qquad F^{a,\mu\nu} = 0~,
\qquad  | \Phi| = 0~,
\ee
which clearly describes the conventional unbroken, or symmetric phase.
In the broken phase the neutral Higgs field gets an expectation value
$\Phi=\inv{\sqrt{2}}(0,\phi_0)^T$ and the weak gauge bosons become massive.
The situation is more transparent when $A^3_\mu$ and $A^Y_\mu$ are
written in terms of the photon field $A_\mu$ and the neutral gauge boson
field $Z_\mu$.  Here we leave out the charged $W$-fields; we will however
discuss their role later.
The equations of motion now become
\bea{fieldeqTree2}
(\partial_\mu - \frac{i\tilde g}{2}Z_\mu)^2\phi_0
 &=& V'(\frac{|\phi_0|^2}{2})\phi_0~,
 \\
\partial_\nu f_Z^{\mu\nu}-\sin \theta_W \partial_\nu f^{\mu\nu}_{\rm ext}
&=&- \frac{i\tilde g}{4}
        \left(\phi_0^*(\partial_\mu- \frac{i\tilde g}{2}Z_\mu)\phi_0
        - \left((\partial_\mu- \frac{i\tilde g}{2}Z_\mu)
        \phi_0\right)^*\phi_0 \right)~,
\\
\label{fAeqom}
\partial_\nu f_A^{\mu\nu} - \cos \theta_W \partial_\nu f^{\mu\nu}_{\rm ext}
&=& 0~,
\eea
where $\tilde g=\sqrt{g^2+g'^2}$. Assume that $\phi_0 = v e^{i\xi(x)}$,
where $v$ is the constant field corresponding to the broken minimum of $V$.
 From \eq{fieldeqTree2} and $V'(v^2/2)=0$ one sees that the $Z$-field
must be a pure gauge: $Z_\mu = (2/\tilde g)\partial_\mu\xi$, and hence
$f_Z^{\mu\nu} = 0$. Note that this is a direct consequence of the
condensation of a field with a $Z$-charge; the Maxwell field does not
couple to $\phi_0$ and therefore is not similarly constrained. Since
$\partial_\nu f^{\mu\nu}_{\rm ext}= 0$, \eq{fAeqom} only tells us that there
are no currents for the Maxwell field strength inside the bubble.  To find
out which value $f_A^{\mu\nu}$ takes we have to minimize Gibbs free energy.
Assuming a pure magnetic field  (electric fields are screened inside the
plasma so ${\bf E = 0}$), we obtain in the broken phase
\be{GibbsEn}
G(\phi,B) = -{\cal L} =V(\phi)
        +\inv2 B^2-\cos\theta_W {\bf B\cdot B}^\ext_{Y}~,
\ee
where ${\bf B}^\ext_{Y}$ is the external classical background hypermagnetic
field and ${\bf B}$ is the Maxwell magnetic field. This is minimized for
${\bf B}=\cos\theta_W {\bf B}^\ext_{Y}$. The Gibbs free energy in the
broken and unbroken phases are thus
\bea{GbGu}
        G_b&=&V(\phi)-\inv2 \cos^2\theta_W (B^\ext_Y)^2~,
        \nn
        G_u&=&V(0)-\inv2 (B^\ext_Y)^2~.
\eea
It is clear that for very large external fields the unbroken solution
(\ref{unbrokenTree}) will be favoured because there the $Z$-component of the
field is nonzero. The critical external field $B^\ext_{Y,crit}$, which marks
the transition from one phase to the other is obtained by setting
$G_u = G_b$, that is
\be{BcritTree}
        \frac{1}{2}\sin^2\theta_W (B^\ext_{Y,crit})^2 =
        V(0)- V(\phi_c)~.
\ee

The magnetic fields strength needed for symmetry restoration in vacuum
would of course be very large. However, close to the critical temperature
the situation is dramatically different, because the free energy
difference between the symmetric and broken minimum can be arbitrarily
small due to the finite temperature corrections to the effective action.
Moreover, since the critical temperature is very sensitive to changes in the
free energies, a relatively small magnetic field can induce a large change in
$\phi_c/T_c$.  For small magnetic fields the most important corrections to the
effective potential are just the usual thermal corrections listed in
Eqs.~($\ref{values}-\ref{ahva}$) above.  With these corrections only, the
analysis above leading to \eq{BcritTree} remains unchanged
except for replacing the potential by the corrected form given by
\eq{veff1loop}.

The polarization effects of the plasma are
typically small in comparision with the tree level terms and they have been
neglected in the discussion so far.
For example, polarization due to one Dirac
fermion with charge $e$ induces a potential term \cite{ElmforsPS94}
\be{Vf}
        \delta V_B^f=
        -\frac{(eB)^2}{12\pi^2}\ln\left(\frac{T}{m_f}\right)~,
\ee
which is a small correction to the tree level energy $B^2/2$, given that
$T^2\gg m_f^2\gg eB$, which is satisfied with a reasonable margin even for
the  relatively strong field with $b\simeq 0.3$.  The potentially most
important effect is caused by polarization of $W$-bosons which become
perturbatively unstable for $eB>M_W^2\phi^2/v^2$.  At one loop
$W$-polarization gives rise to the term (see also \cite{VshitsevZS94})
\bea{newdelta}
\delta_{\rm eff}\phi^3 &\equiv& \frac{M^3_Z\phi^3}{6\pi v^3}
                 + \frac{eB}{2\pi}\left(\sqrt{\frac{M_W^2\phi^2}{v^2}+eB}
                -\sqrt{\frac{M_W^2\phi^2}{v^2}-eB}\right)
        \nn &&
                 - \frac{(2eBT)^{3/2}}{2\pi}\zeta\left(-\frac{1}{2},
                   \frac{1}{2}+\frac{M_W^2\phi^2}{2eBv^2}\right)~,
\eea
where $\zeta(a,x)$ is the generalized zeta-function. In contrast to
the result in \cite{VshitsevZS94} we only include corrections from the
unscreened transverse modes since the longitudinal modes are Debye screened.
The first and the third terms correspond to the terms in \eq{values};
they represent a small correction, smooth in $B$. The  square root terms arise
from the two lowest Landau levels, and the second of these contains the
potentially dangerous instability \cite{JanPoul}.  Its existence signals
the breakdown of perturbation theory due to appearance of a new massless mode.
However, it is known that above the critical field $eB>M_W^2\phi^2/v^2$
a $W$-condensate is formed \cite{JanPoul} and these terms are stabilized
when the theory is expanded around the correct minimum.  In that case the
one-loop corrections will again be found to be small and smooth in $B$.
We conclude that at least for field strenghts below the critical value,
one only need to retain the leading corrections, all of them small,
and not include the square root terms in the computation.
\begin{figure}[t]
\unitlength=1mm
\begin{picture}(100,75)(0,0)
\includegraphics{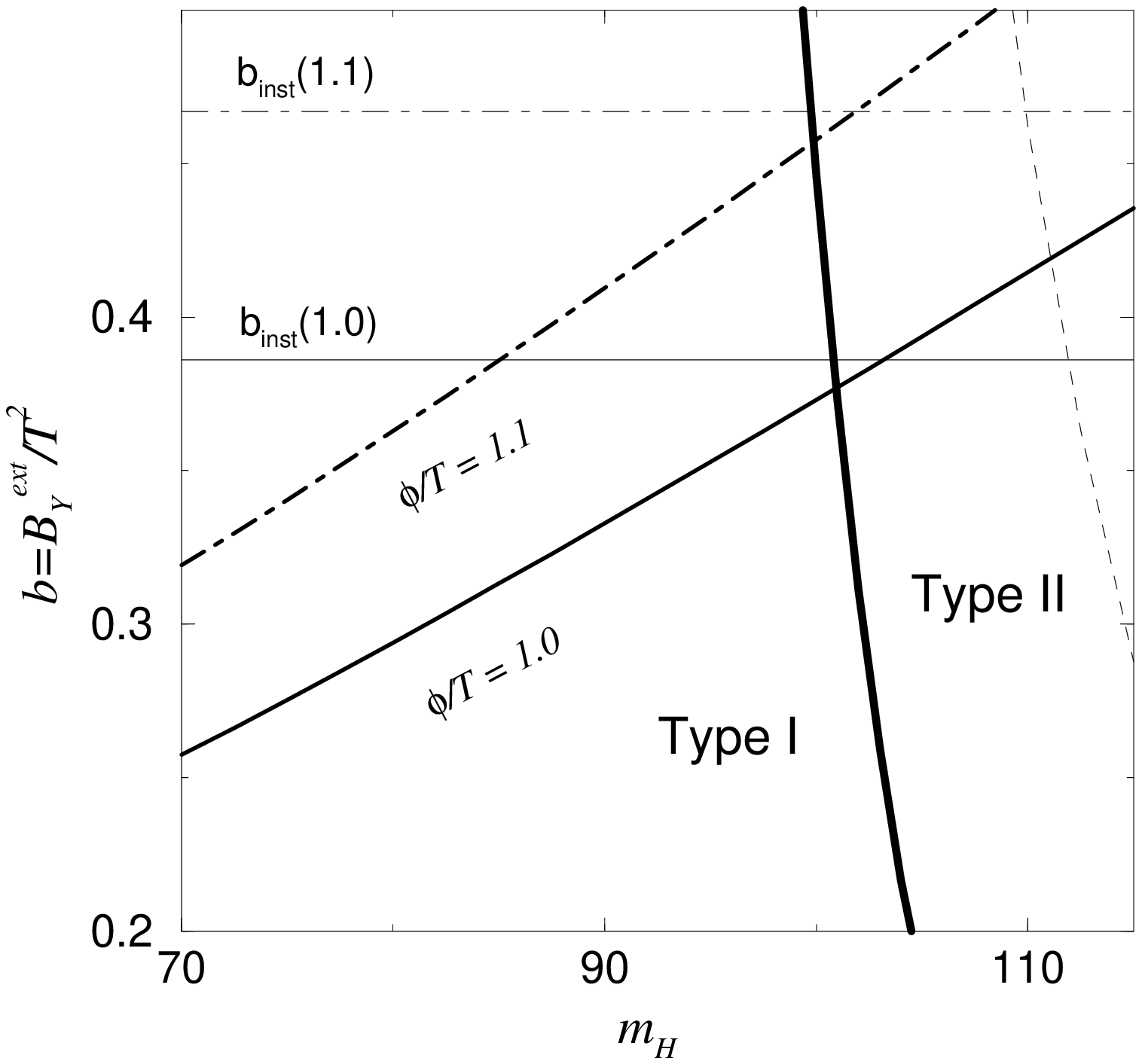}
\end{picture}
\figcap{
The contours of $\phi_c/T_c=1.0$ and $1.1$ at the critical temperature
in the $(B^\ext_Y/T^2, m_H)$-plane and the corresponding field
strengths $b_{\rm inst}$ above which the system is unstable towards formation
of a $W$-condensate.  Shown are also estimates for the boundary of the vortex
phase (type~II), from the argument using correlation lengths (thick solid
line)
and from the instability of the symmetric phase propagator (thin dashed
line).
\label{f:Figure}
}
\end{figure}

We thus obtain the contours shown in \fig{f:Figure} of how strong field is
needed for $\phi_c/T_c$ to equal 1.0 and 1.1.  Note that up to $m_H\simeq
100$ GeV the required field strengths are indeed smaller than the
corresponding critical fields $b_{\rm inst}$ leading to $W$-condensation,
and our analysis is expected to be valid.  Given the constraint
(\ref{sphcons}) we then conclude that a strong enough first order
phase transition for electroweak baryogenesis cannot be ruled out for $m_H
\simleq 100$ GeV, given primordial hypermagnetic field strengths
$B_Y\simgeq (0.3-0.5)T^2$.  Indeed, 
for a Higgs mass of 80~GeV our result agrees
well with \cite{GS}. Above $m_H=100$~GeV our calculations
(nor those of ref. \cite{GS}) cannot be trusted.

In our analysis we have implicitly assumed that MSM corresponds to the
type I superconductor, i.e.\ no vortex solutions arise.  This conclusion is
supported by a crude argument using the correlation lengths of the fields.
Indeed, one expects that the vortex phase can be formed only if the
$Z$-component of the magnetic field can penetrate inside the bubble to a
distance greater than the wall width, given by the inverse of the scalar
mass in the broken phase. That is, a vortex phase (type~II) can form during
the transition only if $m_H(T_c) > m_Z(T_c)$. The border line between the
two phases is displayed in \fig{f:Figure}.
Interestingly, this
phase boundary for the vortex phase is very close to the region of
instability towards $W$-condensation.

Another qualitative criteria for finding the type~II region is to
consider the stability of scalar field configurations against
infrared perturbations in the unbroken phase. During the transition
an instability manifests itself in a tachyonic mode of the scalar
propagator $D^{-1}=\omega^2-k^2-(2n+1)\frac{g'}{2}B_Y-m_H^2(\phi=0,T)$.
Clearly it arises if the critical temperature is lower than the
'vorticity temperature' $T_v$, obtained from
$\frac{g'}{2}bT^2_{v} =m_H^2(\phi=0,T)= -\mu^2 + \alpha T_v^2$.
As shown in \fig{f:Figure}, both this criteria and the one above
lead to qualitatively similar solutions for the boundary between the
type~I and type~II behaviour. They both allow for a sizable region in the
parameter space, up to $m_H\simleq100$ GeV, where our analysis is valid
and which is not constrained by the experiment.  Of course, these are
rough estimates and a numerical analysis of the surface energy
is required to find out the phase diagram accurately.
%
%

In conclusion, the most obvious effect on the electroweak phase transition
from a large scale hyper magnetic backgound field, namely the screening of the
$Z$-component in the broken phase, has been shown to strengthen the phase
transition considerably. A small change in the background field strength makes
a relatively large change in $\phi_c/T_c$ which is the quantity that
determines the sphaleron transition rate.
In order to have a strong enough phase transition to make baryogenesis
possible we need to assume background field strength of $B_Y\simeq 0.3
T^2$.  This is a large field but nevertheless still not ruled out by
observations. Thus we may conclude that
baryogenesis could still be viable within the MSM, provided large primordial
fields existed at the time of EW phase transition and that there is large
enough CP violation.

One should also bear in mind that the actual transition temperature is not
determined by $T_c$ as we assumed above. The action for a finite size bubble
includes the bubble wall and it is necessary to find the bounce action in the
presence of the background field to predict the nucleation rate.
Depending on whether the system acts like a type~I or type~II
superconductor the energy in the wall is either positive or negative.
We have argued that the MSM is a type~I superconductor for the
interesting range of $B$ and
$m_H$; therefore the bubble wall energy is positive which will lower the
actual transition temperature and make it substantially more strongly
first order.

\section* {Acknowledgements}
We thank Poul Olesen for useful remarks on $W$-condensation and
Keijo Kajantie, Misha Shaposnikov and Kari Rummukainen for enlightening
discussions and NorFA for providing partial financial support from the NorFA
grant 96.15.053-O. P.~E. was financially supported by the Swedish Natural
Science Research Council under contract 10542-303, and K.~E. by the Academy
of Finland.
 \vspace{3mm}

%
%

\end{document}